# Crossover in extended Newtonian gravity emerging from thermodynamics

Sumiyoshi Abe [1,2,3] and Peter Ván [4,5,6]

[1] Department of Physics, College of Information Science and Engineering,
 Huaqiao University, Xiamen 361021, China
[2] Institute of Physics, Kazan Federal University, Kazan 420008, Russia
[3] Department of Natural and Mathematical Sciences, Turin Polytechnic University
 in Tashkent, Tashkent 100095, Uzbekistan
[4] Department of Theoretical Physics, Wigner Research Centre for Physics,
 Konkoly-Thege Miklós út 29-33, 1121 Budapest, Hungary
[5] Department of Energy Engineering, Faculty of Mechanical Engineering, Budapest
 University of Technology and Economics, Müegyetem rkp. 3, 1111 Budapest,
 Hungary
[6] Montavid Thermodynamic Research Group, Igmándi út 26, 1112 Budapest, Hungary

Correspondence: suabe@sf6.so-net.ne.jp

**Abstract:** In a recent paper (Ván, P.; Abe, S. *Physica A* **2022**, *588*, 126505), it has been discovered that a scalar field coupled to a fluid and allowed to be a thermodynamic variable in consistency with the second law of thermodynamics is only gravitational and accordingly emergence of extended Newtonian gravity has been predicted. The resulting field equation for the potential of this emergent force is nonlinear and admits the logarithmic potential as a singular solution, suggesting its relevance to the dark matter conundrum. Here, a general analysis of the nonlinear field equation is performed. It is found that the emergent force field exhibits the unsharp crossover between the $1/r$ and $1/r^2$ forces outside the fluid, depending on a spatial scale characteristic of the present theory to be observationally tested in the context of the dark matter conundrum. Then, the action functional is constructed for the potential of the emergent field and the field energy is shown to be free from an infrared divergence. A comment is also made on the difference of the present theory from MOND (modified Newtonian dynamics).

**Keywords:** extended Newtonian gravity; thermodynamics; crossover of the emergent force; action functional; dark matter conundrum

# 1. Introduction

Among the forces in nature, gravity is outstanding in the sense that it does not depend on the details of compositions/structures of matters. Such a universal feature may remind one of thermodynamics. This point suggesting a possible relevance between thermodynamics and gravity has been investigated during the last decade [1-3] (see also Reference [4] and the works quoted therein as well as the earlier thoughts and investigations in References [5-7]). In particular, widely noted is the proposal in Reference [1] that gravity may be an entropic force.

In a recent work [8], we have addressed ourselves to the following radical question. Consider a scalar field potential coupled to a fluid in a nonequilibrium state. Then, is it possible to treat the whole system in consistence with the laws of thermodynamics? The answer to this question has turned out to be remarkable: *the field, which can be a thermodynamic variable, is only gravitational*. There, gravity is in fact outstanding and unique because the energy (density) of the field potential has been shown necessarily to be negative [9] for the consistency. It is noted that this result, albeit nonrelativistic, has been derived without additional requirements such as the Unruh effect [10] and the holographic principle [11]. Guided by this fact, we have studied the relaxation process of the whole system by employing the standard Onsager-Casimir theory [12] for the entropy production during the process, i.e. the second law of thermodynamics. We have discovered that the field may relax to the one, which includes an extension of Newtonian gravity: emergence of extended Newtonian gravity from thermodynamics. We have considered a spatial scale. It is typically the size of a galaxy given in terms of its luminous region (such as the Petrosian radius), up to which the theory is described by Newtonian gravity, whereas the emergent non-Newtonian becomes relevant in the distance $r$ larger than that, where $r$ is the distance from a representative point in a galaxy as a finite fluid distribution. Specifically, in terms of the magnitude of the force field felt by a test particle outside the fluid, the $1/r$ force appears as $r$ increases.

The field equation for the potential of emergent gravity derived in Reference [8] is nonlinear and admits the logarithmic solution leading to the $1/r$ force mentioned above. This might apparently possesses an infrared divergence of the field energy. In this paper, we show that this is not the case. We impose the spherical symmetry on that nonlinear field equation and see that it becomes an equation of the Bernoulli-Riccati type. Analyzing it, we find that a solution of physical interest shows existence of a spatial scale. We show that, in terms of the scale, there exists the crossover between the $1/r$ and Newtonian $1/r^2$ forces (i.e. the logarithmic potential and ordinary $1/r$ potential, respectively) outside the fluid. We construct the action functional for the emergent gravitational field potential and prove that the field energy is free from an infrared divergence even for the pure logarithmic potential. We discuss the implication of the emergent $1/r$ force outside the fluid to the dark matter conundrum. Finally, we make a comment on how this theory is different from modified Newtonian dynamics commonly abbreviated as MOND [13] (see also References [14,15] and the works quoted therein).

## 2. Nonlinear Field Equation for Extended Newtonian Gravity Emerging from Thermodynamics

As shown in Reference [8], the scalar field potential $\phi$ up to the galaxy scale may obey the following equation:

$$\frac{\partial \phi}{\partial t} = \frac{l^2}{\tau}\left(\nabla^2 \phi - 4\pi G \rho\right), \tag{1}$$

where $G$ is Newton's gravitational constant and $\rho$ is the mass density of a fluid. $l$ is relevant to typical size of microscopic spatial variations of the whole fluid-field system. Equation (1) describes the relaxation to a "stationary state" (formation of a galaxy, presumably) with the relaxation time $\tau$. Thus, time dependence of $\phi$ has its origin in the nonequilibrium nature of the fluid. Since the second law of thermodynamics is of a central issue here, time-reversal invariance is violated in Equation (1). In the limit of relaxation, the Poisson equation

$$\nabla^2 \phi = 4\pi G \rho \tag{2}$$

is shown to hold. Therefore, inside the fluid distribution, i.e. presumably up to the galaxy scale, the potential relaxes to that of Newtonian gravity. This field is said to be "elementary". In fact, there are no *a priori* reasons to discard Newtonian gravity in the nonrelativistic regime.

On the other hand, there appears another field that extends Newtonian gravity. We call it emergent gravity ("emergent" in contrast to "elementary" mentioned above). Since the emergent degrees of freedom are generally considered to be relevant on a larger scale, it is natural to assume that the extended Newtonian gravity may play a role outside the fluid distribution, or a galaxy. In the relaxation limit, the stationary field equation in vacuum is shown to be given as follows [8]:

$$\nabla^2 \phi = K (\nabla \phi)^2 \qquad (3)$$

with

$$K = \frac{L_{12}}{24\pi G \det(L)}, \qquad (4)$$

where $L$ is the matrix of the kinetic coefficients $L_{ij}$'s ($i, j = 1, 2$) of the coupled field-fluid system satisfying the Onsager-Casimir reciprocity [12]. In the particular case when $K = 0$, Equation (3) is reduced to Equation (2) without the source term. This emergent part of gravity may in fact be dominant in the region outside a fluid since Equation (3) predicts the logarithmic potential as the singular solution, i.e. the solution independent of boundary conditions, and therefore the force behaves as $1/r$ (see the next section). It is however noted that $K$ should be positive if such an emergent force is required to be attractive [8]. So far, there are no *a priori* reasons known for ruling out negative $K$.

**3. Crossover in Emergent Theory of Extended Newtonian Gravity**

Now, we come to the main part of the present work. Our discussion is based on a full analysis of Equation (3) describing the emergent gravity outside a fluid under the spherical symmetry.

In the spherically-symmetric case, Equation (3) for $\phi(r)$ becomes as follows:

$$\frac{1}{r^2}\left(r^2\frac{d^2}{dr^2}+2r\frac{d}{dr}\right)\phi = K\left(\frac{d\phi}{dr}\right)^2. \tag{5}$$

Let us set

$$f = -\frac{d\phi}{dr}, \tag{6}$$

which is the one and only component (i.e. the radial component) of the force field felt by a test particle per its mass. Then, $f$ satisfies

$$\frac{df}{dr} = -\frac{2}{r}f - Kf^2. \tag{7}$$

This belongs to the class of equations of the Bernoulli-Riccati type. Hence, first let us find a singular solution independent of an arbitrary constant. In the present case, it is given by

$$f_S = -\frac{1}{Kr}. \tag{8}$$

Then, substituting

$$f = \tilde{f} + f_S \tag{9}$$

into Equation (7), we obtain

$$\frac{d\tilde{f}}{dr} = -K\tilde{f}^2, \tag{10}$$

in which separation of variables is seen to be realized. Thus, we have the following two solutions:

$$\tilde{f} = 0, \tag{11}$$

$$\tilde{f} = \frac{1}{Kr+c}, \tag{12}$$

where $c$ in Equation (12) is a constant. Equation (11), which is the trivial solution of Equation (10) yielding the $1/r$ force, corresponds to the limiting case ($c \to \pm\infty$) of the nontrivial solution in Equation (12) with fixed $r$.

If Equation (12) is used in Equation (9), then the following force is obtained:

$$f = -\frac{c}{(Kr)^2 + cKr}. \tag{13}$$

It is natural to require that the force does not have a singularity at a finite value of $r$. Therefore, $K$ and $c$ must have the same sign. Accordingly, the force is attractive (repulsive) if both $K$ and $c$ are positive (negative), and henceforth the case of nonnegative $K$ will be considered. In the particular case when $c = 0$, extended Newtonian gravity does not emerge.

The solution in Equation (13) indicates that there exists a nontrivial spatial scale:

$$R \equiv c/K, \qquad (14)$$

over which the $1/r^2$ behavior becomes dominant. For $r \gg R$, the system loses its fluid nature and comes to behave pointlike. Therefore, gravity should return to be Newtonian. Since $f \sim -c/(Kr)^2$, we have $c = GMK^2$, where $M$ is the total mass of the fluid. Accordingly, the scale in Equation (14) is rewritten as

$$R = GMK. \qquad (15)$$

The case of particular interest appears between the size of the fluid distribution and $R$, in which the exotic emergent $1/r$ force is relevant. The theory predicts the crossover:

$$1/r \rightarrow 1/r^2 \qquad (16)$$

as $r$ increases. In Figure 1, a plot of the emergent force is presented for visualization of this feature.

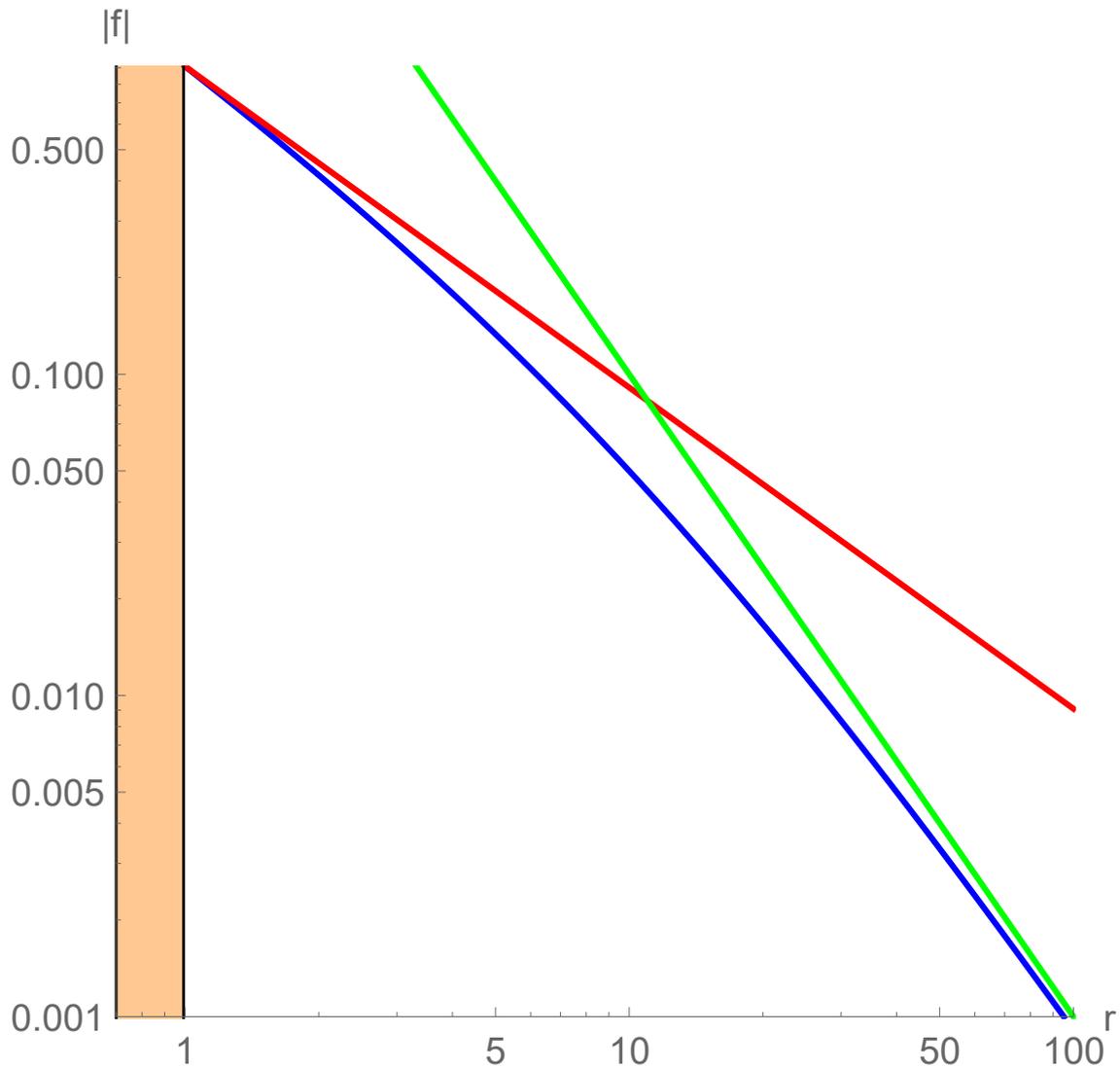

**Figure 1.** The log-log plot of Equation (13). The orange region $r \leq 1$ depicts the distribution of a fluid, in which Newtonian gravity is unchanged. However, the explicit form of the force felt by a test particle depends on the distribution, and therefore its plot is not presented, here. [For example, in the case of the spherical and uniform mass distribution, the solution of Equation (2) governing Newtonian gravity is quadratic in $r$ and the force felt by a test particle inside the distribution is linear, not $1/r^2$.] The emergent force is shown by the blue curve in the region $r > 1$. $K$ and $c$ are taken to be 1

and 10, respectively, and accordingly the scale $R$ in Equation (14) is $R=10$. The red line shows the $1/r$ force, whereas the $1/r^2$ force is indicated by the green line. The unsharp crossover occurs around $r \sim R = 10$. All quantities are dimensionless.

When $K$ vanishes, the scale $R$ in Equation (15) collapses to the fluid size and no extended gravity emerges.

Now, if the fluid is regarded as a galaxy, then the emergent $1/r$ force explains the observed flat rotation curve without dark matter. In Section 5, we come back to this issue and make a comment on Equation (15) in view of the dark matter conundrum.

## 4. Action Functional and Field Energy of Emergent Gravity

In the present theory, extended Newtonian gravity described by the field equation (3) is the one that emerges from thermodynamics in the limit of relaxation of the total field-fluid system to a stationary state of it. Although the second law of thermodynamics does not have a good affinity with the action principle due to its irreversible nature, the irreversibility may not matter after the relaxation. Here, we discuss this issue and consequently define the field energy of emergent gravity.

The action functional we present here reads

$$S = \int dx^4 \, \mathcal{L} \tag{17}$$

with the Lagrangian density

$$\mathcal{L} = \frac{1}{2} e^{-2K\phi} \, \partial_\mu \phi \, \partial^\mu \phi, \qquad (18)$$

where the 4-dimensional sign convention $(-,+,+,+)$ is used. The original field $\phi$ will be kept although the field redefinition $\phi \to \Phi \propto e^{-K\phi}$ simplifies the action. Time dependence here describes possible propagation of the field and *should not be confused with that in the relaxation process* as in Equation (1). Variation of the action with respect to $\phi$ yields the following field equation:

$$\partial_\mu \partial^\mu \phi = K \, \partial_\mu \phi \, \partial^\mu \phi. \qquad (19)$$

From this, the field equation (3) is obtained in the time-independent case.

Since $\phi$ is a potential, it is defined up to an arbitrary additive constant. Let the constant be denoted by $\phi_0$. In fact, the transformation $\phi \to \phi + \phi_0$ does not change the field equation for $\phi$. The action in Equation (17), however, is not invariant under such a transformation: $S \to S/\lambda^2$, where $\lambda$ is a positive constant given by $\lambda = e^{K\phi_0}$. Therefore, in order for the action to be invariant, this transformation has to be combined with the rescaling of the spacetime coordinate: $x^\mu \to \lambda x^\mu$. It should be noticed that such invariance cannot be realized in 2-dimentional spacetime.

The canonical momentum density conjugate to $\phi$ is $\pi = \partial \mathcal{L}/(\partial \phi/\partial t) = -e^{-2K\phi} \partial \phi/\partial t$ and the Hamiltonian density $\mathcal{H} = \pi \, \partial \phi/\partial t - \mathcal{L}$ is calculated to be

$$\mathcal{H} = -\frac{1}{2}e^{2K\phi}\pi^2 - \frac{1}{2}e^{-2K\phi}(\nabla\phi)^2. \tag{20}$$

Therefore, the energy of the time-independent emergent field is given by

$$E_{ef} = -\frac{1}{2}\int dx^3\, e^{-2K\phi}(\nabla\phi)^2. \tag{21}$$

In the limit $K \to 0$, Equation (3) becomes reduced to the ordinary Poisson equation in vacuum i.e. the Laplace equation, and Equation (21) reproduces the negative field energy of Newtonian gravity [9].

As discussed in the preceding section, there are two distinct cases, in which $\phi$ behaves differently in the limit $r \to \infty$. One is of the crossover in Equation (16) where $\phi$ asymptotically decreases as $1/r$ like in Newtonian gravity. Therefore, the field energy in Equation (21) does not have an infrared divergence in this case. On the other hand, the case of the $1/r$ force in Equation (8) corresponding to Equation (11) is intriguing. There, the potential is logarithmic: $\phi = (\ln r)/K$, up to an additive constant. Again in this case, the field energy is seen to have no infrared divergence because of the factor $e^{-2K\phi}$ playing a crucial role. Consequently, the field energy in the theory is free from an infrared divergence.

## 5. Conclusion and Remark on Dark matter Conundrum

We have performed a detailed analysis of the nonlinear field equation describing extended Newtonian gravity emerging from thermodynamics by imposing the spherical symmetry. We have shown that, depending on the kinetic coefficients of the coupled field-fluid system relaxed to a stationary state and boundary conditions on the emergent field, the theory exhibits nontrivial behaviors: crossover of the force, repulsive force and even vanishing emergent gravity. We have also constructed the action for emergent gravity, in which we have *apparently* implemented the Lorentz symmetry to describe possible propagation of the emergent field. It is however noted that in general relativity $\phi$ is not a scalar quantity. This fact seems make it highly nontrivial to endow the theory with general covariance. We wish to discuss this issue elsewhere.

In the intermediate region between the size of a galaxy and $R$ in Equation (15), the $1/(Kr)$ force can emerge. This force gives rise to the constant magnitude $V_*$ of the rotational velocity outside a galaxy

$$V_* = 1/\sqrt{K} . \tag{22}$$

If the theory is applied to the problem of the constant rotational speeds (i.e. the flat rotation curve) without assuming existence of dark matter, then from the observational data in References [16,17] $K$ may be estimated to be in the range between $1.0 \times 10^{-11} s^2/m^2$ and $1.0 \times 10^{-8} s^2/m^2$. Thus, Equation (15) predicts the relation between the scale $R$ and the mass of each galaxy. In addition, according to the recent observational results [18,19], there certainly exist galaxies that do not contain dark matter. In the present theory, they may correspond to the cases where $K$ is negligibly small.

In recent years, a lot of efforts have been made on MOND, modified Newtonian dynamics [13-15]. MOND changes Newtonian gravity by introducing a constant having the dimension of acceleration in order to explain the observed constant rotation speeds outside the visible regions of galaxies [16,17] without assuming existence of dark matter. In this context, the present theory is radically different from MOND. At the elementary level, we do not change Newtonian gravity characterized by *G*. Extended Newtonian gravity characterized by *K* emerges from thermodynamics between the scale of a galaxy and *R* given in terms of *K* and the galaxy mass *M* as in Equation (15). And, *K* comes from the fluid nature of each galaxy.

**Author Contributions:** The problem has been formulated through the discussions between S.A. and P.V. and both of them have equally contributed to the work. S.A. has organized the paper and P.V. has agreed to publish it.

**Funding:** The work of S.A. has been supported by the Program of Fujian Province and the Program of Competitive Growth of Kazan Federal University from the Ministry of Education and Science of the Russian Federation. P.V. has been supported by the grants: National Research, Development and Innovation Office–NKFIH 123815, TUDFO/51757/2019-ITM (Thematic Excellence Program) and FIEK-16-1-2016-0007, as well as the Higher Education Excellence Program of the Ministry of Human Capacities in the frame of Nanotechnology research area of Budapest University of Technology and Economics (BME FIKP-NANO).

**Institutional Review Board Statement:** Not applicable.

**Informed Consent Statement:** Not applicable.

**Data Availability Statement:** Not applicable.

**Conflicts of Interest:** Not applicable.